\documentclass[12pt]{article}
\usepackage{amssymb}
\usepackage{lscape}
\usepackage{tabls}
\def\beq{\begin{equation}}
\def\eeq{\end{equation}}
\def\be{\begin{eqnarray}}
\def\ee{\end{eqnarray}}
\def\nn{\nonumber}
\def\lra{\longrightarrow}
\def\lb{\left(}
\def\rb{\right)}
\def\5{[5]_{q^{-2}}!}
\def\qL{$q$-Laguerre}
\begin{document}
\begin{center}
{\LARGE\sf New connection formulae for the $q$-orthogonal polynomials 
via a series expansion of the $q$-exponential}

\smallskip
\smallskip

R. Chakrabarti$^{a}$, R. Jagannathan$^{b}$ and  S. S. Naina 
Mohammed$^{a}$

\smallskip

{\small {\it $^{a}$ Department of Theoretical Physics, University of 
Madras \\ 
Guindy Campus, Chennai 600025, India}}

\smallskip

{\small {\it $^{b}$ The Institute of Mathematical Sciences \\  
C.I.T. Campus, Tharamani, Chennai 600113, India}}
\end{center}
 
\begin{abstract}
Using a realization of the $q$-exponential function as an infinite 
multiplicative series of the ordinary exponential functions we obtain 
new nonlinear connection formulae of the $q$-orthogonal polynomials 
such as $q$-Hermite, $q$-Laguerre and  $q$-Gegenbauer polynomials in 
terms of their respective classical analogs.
\end{abstract}
\section{Introduction}
\label{Intro}
\renewcommand{\theequation}{\arabic{section}.{\arabic{equation}}}
\setcounter{equation}{0}
With the advent of quantum groups $q$-orthogonal polynomials are 
objects of special interest in both mathematics and physics.  For 
instance, $q$-deformed harmonic oscillator provides a group-theoretic 
setting for the $q$-Hermite and the $q$-Laguerre polynomials.  In the 
mathematical framework needed to describe the properties of these 
$q$-polynomials, such as the recurrence relations, generating functions, 
and orthogonality relations, Jackson's $q$-exponential plays a key role.  
Recently, Quesne~\cite{Q04} has expressed Jackson's $q$-exponential as a
multiplicative series of the ordinary exponentials with known 
coefficients in closed form.  In this scenario it becomes imperative to 
investigate the effects of this result on the theory of $q$-orthogonal 
polynomials. The present work is an attempt in that direction. In 
particular, we employ the said result to obtain new nonlinear connection 
formulae for $q$-Hermite, $q$-Laguerre and $q$-Gegenbauer polynomials in 
terms of their respective classical counterparts.   

Jackson actually introduced two related $q$-exponentials:
\be
e_q(z) & = & \sum_{n=0}^\infty \frac{1}{(q;q)_n} z^n,\nonumber \\
E_q(z) & = & \sum_{n=0}^\infty \frac{q^{n(n-1)/2}}{(q;q)_n} z^n, 
\label{jack_exp}
\ee 
where 
\be
&  & (a;q)_0 = 1, \quad 
     (a;q)_n = \prod_{k=0}^{n-1} (1-aq^k), \quad
     (a;q)_{\infty} = \prod_{k=0}^{\infty} (1-aq^k). 
\ee  
The classical limits of the above $q$-exponentials read
\be
\lim_{q \lra 1} e_q((1-q)z) & = & \exp (z), \qquad 
\lim_{q \lra 1} E_q((1-q)z) = \exp (z).
\ee  
Heine's $q$-binomial theorem provides \cite{GR90} the following 
multiplicative series: 
\be
e_q(z) & = & \frac{1}{(z;q)_\infty}, \qquad 
E_q(z) = (-z;q)_\infty, 
\ee 
such that 
\be
&  & e_q(z)E_q(-z) = 1. 
\label{eE}
\ee
In physics literature another form of $q$-exponential commonly occurs.  
This is given by 
\be
\exp_q(z) & = & \sum_{n=0}^\infty \frac{1}{[n]_q!}z^n, 
\label{phys_qexp}
\ee
where 
\beq
[n]_q = \frac{1-q^n}{1-q},\quad [n]_q! = [n]_q[n-1]_q\cdots [1]_q, 
\quad [0]_q! = 1. 
\label{q_num}
\eeq
Comparing the first equation in (\ref{jack_exp}) and (\ref{phys_qexp}) 
it is evident that 
\be
\exp_q(z) & = & e_q((1-q)z). 
\ee 
Quesne~\cite{Q04} expressed the $q$-exponential (\ref{phys_qexp}) as 
a product series of the ordinary exponentials as follows: 
\be
\exp_q(z) & = & \exp\lb \sum_{k \in \mathbb{N}} c_k(q)z^k \rb, \quad 
c_k(q) = \frac{(1-q)^{k-1}}{k[k]_q}.   
\label{Ques_prod}
\ee  
This expansion allows us to write
\be
e_q(z) & = & \exp\lb \sum_{k \in \mathbb{N}} \frac{z^k}{k(1-q^k)} \rb, 
\label{eqseries}
\ee
and 
\be
E_q(z) & = & \exp\lb \sum_{k \in \mathbb{N}}
                     \frac{(-1)^{k+1}z^k}{k(1-q^k)} \rb. 
\label{Eqseries}
\ee
These results are particularly suitable for applications in the theory of 
special functions. Specifically, they may be readily employed to develop a 
general procedure for obtaining connection formulae for various $q$-orthogonal 
polynomials. First, the generating function of a suitable $q$-orthogonal 
polynomial may be expressed as an infinite product of generating functions 
of the corresponding classical polynomial.  This, in turn, may be utilized 
to develop a Diaphontine partition equation.  As successive terms in the 
said product series have increasing exponents of the arguments, the partition 
equation has a finite number of solutions, and each $q$-orthogonal polynomial 
can be expressed in terms of its classical counterparts of equal or lesser 
orders. In the following sections we illustrate this procedure in the 
cases of $q$-Hermite, $q$-Laguerre and $q$-Gegenbauer polynomials.

\section{$q$-Hermite polynomials}
\label{q_Hermite}
\setcounter{equation}{0}
It has been observed \cite{HW99} that the $q$-Hermite polynomials are
components of the eigenfunctions of the Hamiltonian of the $q$-deformed 
harmonic oscillator. These authors investigated the recurrence relation,
the generating function, the orthogonality relation and other 
properties of the said polynomials. The generating function of the 
$q$-Hermite polynomials has been expressed \cite{HW99} in terms of the 
$q$-exponentials:   
\be
{\sf G}_{q}(z; t) &\equiv& E_{q^{-2}}(2 (1 - q^{-2})\, z t)\,\,
e_{q^{-4}}\Big(- \frac{2\, (1-q^{-4})\, t^{2}}{q\,(1 + q^{-2})} \Big)\nn\\
&=& \sum_{n = 0}^{\infty}\,\frac{q^{n/2}}{[n]_{q^{-2}}!}\,
H_{n}(z; q)\,t^{n}.
\label{q_Her_gen}
\ee   
In the $q \rightarrow 1$ limit, above function ${\sf G}_{q}(z; t)$ 
yields the well-known  generating function of the classical Hermite 
polynomials:
\beq
{\sf G}(z; t) \equiv \exp (2 z t - t^{2}) = \sum_{n = 0}^{\infty}\, 
\frac{1}{n!}\, H_{n}(z)\, t^{n}.
\label{c_Her_gen}
\eeq
For later use in our expansion scheme the classical Hermite polynomials
are listed below:
\beq
H_{n}(z) = \sum_{\ell = 0}^{[n/2]} (-1)^{\ell}\,
\frac{n!\,(2 z)^{n - 2 \ell}}{\ell!\, (n - 2 \ell)!},
\label{cla_Her}
\eeq
where the symbol $[n/2]$ means the largest integer smaller or equal to 
$n/2$. Following (\ref{eqseries}) and (\ref{Eqseries}) we recast the 
$q$-exponentials in (\ref{q_Her_gen}) as 
\be
E_{q^{-2}}(2 (1 - q^{-2})\, z t) &=&
\exp \Big(\sum_{k \in {\mathbb N}} (-1)^{k+1}\,c_{k}(q^{-2})\,
(2 z t)^{k}\Big),\\
e_{q^{-4}}\Big(- \frac{2\, (1 - q^{-4})\, t^{2}}{q\,(1 + q^{-2})} \Big) 
&=&\exp \Big(\sum_{k \in {\mathbb N}} (-1)^{k}\,c_{k}(q^{-4})\,
\Big(\frac{2 t^{2}}{q (1 + q^{- 2})}\Big)^{k}\Big). \nonumber
\label{h_exp}
\ee
Introducing the following parameters
\be
\tau_{k} &=& (- 1)^{(k + 1)/2}\,
\Big(\frac{2}{q\, [2]_{q^{-2}}}\Big)^{k/2}\,\sqrt{c_{k}(q^{-4})}
\,\,t^{k},\nn\\
\zeta_{k} &=& (- 1)^{(k + 1)/2}\,
(2\, q\, [2]_{q^{-2}})^{k/2}\,
\frac{c_{k}(q^{-2})}{2\, \sqrt{c_{k}(q^{-4})}}\, z^{k},
\label{h_parameter}
\ee
the generating function (\ref{q_Her_gen}) may be expressed as a 
multiplicative series of the classical generating functions:
\beq
{\sf G}_{q}(z; t) = \prod_{k \in {\mathbb N}}\,
{\sf G}(\zeta_{k}; \tau_{k}).
\label{h_gen_exp}
\eeq
Employing the expansions (\ref{q_Her_gen}) and (\ref{c_Her_gen}) of the
above generating functions, and comparing coefficients of equal power of 
$t$ on both sides, we obtain a connection formula for the $q$-Hermite
polynomials in terms of its classical partners of lower dimensions:   
\be
\frac{q^{n/2}}{[n]_{q^{-2}}!}\,H_{n}(z; q) &=& 
\Big(\frac{2}{q [2]_{q^{-2}}}\Big)^{n/2}\,\, 
\sum_{n_{1}, n_{2}, \cdots
\atop = \,0}^{\infty} (- 1)^{(n + \sum_{k \in {\mathbb N}} n_{k})/2}\,
\quad \times \nn\\ 
& & \times \,\,\,\prod_{k \in {\mathbb N}}
\Big[(c_{k}(q^{- 4}))^{n_{k}/2}\,
\frac{H_{n_{k}}(\zeta_{k})}{n_{k}!}\Big]\;
\delta_{\sum_{k \in {\mathbb N}} k n_{k}, n}.
\label{q_Herm_prod}
\ee
The solutions of Diaphontine partition relation
\beq
\sum_{k \in {\mathbb N}} k n_{k} = n
\label{Her_part}
\eeq
select the product structure of the classical classical Hermite 
polynomials appearing on the rhs of 
(\ref{q_Herm_prod}). As an illustration we here use the connection 
formula (\ref{q_Herm_prod}) for reconstructing the $q$-Hermite 
polynomial $H_{5}(z; q)$. The solution of the partition equation 
(\ref{Her_part}) and the corresponding contributions to the connection 
formula (\ref{q_Herm_prod}) are listed in Table \ref{H_T}. Combining 
the entries of the second column we may easily reproduce the 
well-known \cite{HW99} result for $H_{5}(z; q)$:  
\beq
H_{5}(z;q) = 32 q^{- 45/2}\, z^{5} - 16 q^{-19/2}\,[2]_{q^{-4}}\,
[5]_{q^{-2}}\,z^{3} + 8 q^{- 9/2}\, [3]_{q^{-2}}\,[5]_{q^{-2}}\, z.
\label{Hq_5}
\eeq
\section{$q$-Laguerre polynomials}
\label{q_Laguerre}
\renewcommand{\theequation}{\arabic{section}.{\arabic{equation}}}
\setcounter{equation}{0}
The theory of the \qL\,  polynomials has been studied \cite{M81, FV91} 
extensively. They appear \cite{FV91} in the representation theory of the 
enveloping algebra of the $q$-deformed Heisenberg algebra. These authors 
observed \cite{FV91} that the generating function of \qL \, 
polynomials may be cast in the form  
\beq
{\cal G}^{(n)}_{q}(z;t) \equiv
E_{q}(-(1-q)\, z t)\,\Big(-\frac{q}{t}; \,q\Big)_{n}\, t^{n} =
\sum_{k} q^{(n - k)\,(n - k + 1)/ 2}\,L^{(n-k)}_{k}(z; q)\,t^{k},
\label{qL_gen}
\eeq
which may be viewed as the $q$-analog of the generating function    
\cite{E53} of the classical Laguerre polynomials: 
\beq
{\cal G}^{(n)}(z;t) \equiv \exp (- z t)\,(1 + t)^{n} = 
\sum_{k} L^{(n - k)}_{k}(z)\, t^{k}.
\label{L_gen}
\eeq
For the purpose of later use we list here the the classical Laguerre 
polynomials as
\beq
L^{(n - k)}_{k}(z) = \sum_{\ell = 0}^{k} (- 1)^{\ell}
\Big( \begin{array}{c}n\\k - \ell\end{array} \Big)\,\frac{z^{\ell}}
{\ell !}. 
\label{L_list}
\eeq

\par

The product structure (\ref{Eqseries}) may now be utilized to express the 
$q$-Laguerre generating function (\ref{qL_gen}) as a multiplicative 
series of the classical Laguerre generating function (\ref{L_gen}):
\beq
{\cal G}^{(n)}_{q}(z;t) = \prod_{k \in \mathbb{N}} 
\Big[{\cal G}^{(n_{k})}(c_{k}(q) z^{k};t^{k})\,
(1 + t^{k})^{- n_{k}}\Big]\,
\Big(-\frac{q}{t}; \,q\Big)_{n}\, t^{n},
\label{gen_expn}
\eeq
where any integer set $\{n_{k}\}$ may be used in the rhs. Each set of 
auxiliary integer parameters $\{n_{k}\}$ provides an expansion 
scheme for the $q$-Laguerre polynomials. The deformed $q$-Laguerre 
polynomials, when reconstructed via our expansion scheme involving the 
classical Laguerre polynomials, must not depend on the intermediate 
auxiliary integer parameters $\{n_{k}\}$. As observed below, precisely 
this happens. The expansion scheme (\ref{gen_expn}) forms the key 
ingredient of our method. To explicitly obtain the connection formula 
for the \qL\, polynomials we proceed by expressing the $q$-binomial 
theorem \cite{GR90} as 
\beq
\Big(-\frac{q}{t}; \,q\Big)_{n}\, t^{n} = 
\sum_{\ell = 0}^{n}\,q^{(n - \ell) (n - \ell + 1)/2}\,
\Big[\begin{array}{c}n\\{\ell}\end{array}\Big]_{q}\,t^{\ell}, 
\label{q_bino}
\eeq
where $\Big[\begin{array}{c}n\\{\ell}\end{array}\Big]_{q} =
\frac{[n]_{q}!}{[\ell]_{q}!\;[n - \ell]_{q}!}$. Using Pochammer
symbol 
\beq
(\alpha)_{\ell} = \prod^{\ell-1}_{j=0} (\alpha + j),\quad 
(\alpha)_{0} = 1,
\label{poch}
\eeq
we rewrite
\beq
(1 + t)^{- n} = \sum_{\ell = 0}^{\infty} (- 1)^{\ell}
\frac{(n)_{\ell}}{\ell !}.
\label{cl_bino}
\eeq
Systematic use of the results (\ref{L_gen}), (\ref{q_bino}), and 
(\ref{cl_bino}) on the rhs of the expansion scheme 
(\ref{gen_expn}) now yields the relation
\be
{\cal G}^{(n)}_{q}(z;t) &=&  
\sum_{\ell = 0}^{n} \sum_{\{k_{j}\}} \sum_{\{\ell_{j}\}} 
\prod_{j \in \mathbb{N}} \Big( (-1)^{\ell_{j}}\, 
\frac{(n_{j})_{\ell_{j}}}{\ell_{j}!}\,
L^{(n_{j} - k_{j})}_{k_{j}}(c_{j}(q) z^{j})\, t^{j (k_{j} + \ell_{j})}
\Big) \quad \times \nn\\
& & \times\,\,\, q^{(n - \ell) (n - \ell + 1)/2}\,
\Big[\begin{array}{c}n\\{\ell}\end{array}\Big]_{q}\,t^{\ell}.
\label{L_exp_pro}
\ee
Using the second equality in (\ref{qL_gen}) and comparing the 
coefficients of $t^{k}$ on both sides of (\ref{L_exp_pro}), we obtain
the promised nonlinear connection formula
\be
&&q^{(n - k) (n - k + 1)/2} L^{(n - k)}_{k}(z;q)\nn\\ 
&&= \sum_{\ell = 0}^{n} \sum_{\{k_{j}\}} \sum_{\{\ell_{j}\}} 
\prod_{j \in \mathbb{N}} 
\Big( (-1)^{\ell_{j}}\, \frac{(n_{j})_{\ell_{j}}}{\ell_{j}!}\,
L^{(n_{j} - k_{j})}_{k_{j}}(c_{j}(q) z^{j})\Big) \quad \times \nn\\
&& \times\,\,\, q^{(n - \ell) (n - \ell + 1)/2}\,
\Big[\begin{array}{c}n\\{\ell}\end{array}\Big]_{q}\,
\delta_{\sum_{j} j (k_{j} + \ell_{j}) + \ell,\, k}.
\label{L_exp_fn}
\ee
The solutions of the Diaphontine partition relation
\beq
\sum_{j \in \mathbb{N}} j (k_{j} + \ell_{j}) + \ell = k
\label{part_eq}
\eeq
determine the set of classical Laguerre polynomials contributing to the 
expansion of a particular \qL\, polynomial. In continuation of the 
discussion following (\ref{gen_expn}) we note that the lhs of 
(\ref{L_exp_fn}) is independent of the set $\{n_{j}\}$;
and, consequently, each set of allowed $\{n_{j}\}$ provide an expansion of 
the \qL\, polynomial $L^{(n - k)}_{k}(z;q)$. This is a generic feature of 
our procedure. Equations (\ref{L_exp_fn}) and (\ref{part_eq}) form the 
main results of this section. To illustrate our process with an example 
we here enumerate the contributing terms for $n = 3, k = 3$. The 
corresponding solutions of the partition relation (\ref{part_eq}), and their
respective contributions to $L_{3}^{(0)}(z;q)$ are listed in 
Table \ref{L_T}. The entries on the first column refer to the non-zero 
elements in the solutions of the partition equation (\ref{part_eq}) 
for $k = 3$. Using the classical Laguerre polynomials given in 
(\ref{L_list}), we may now directly recover the polynomial 
$L_{3}^{(0)}(z;q)$ by summing the entries on the second column of 
Table \ref{L_T}. As noted earlier, the dependences on the auxiliary 
integer parameters $n_{1}, n_{2}, n_{3}$ in the contributions to 
$L_{3}^{(0)}(z;q)$, when summed, disappears. Summing the contributions 
displayed on the second column in Table \ref{L_T} we obtain
\beq
L_{3}^{(0)}(z;q) = 1 - q \Big[\begin{array}{c}3\\1\end{array}\Big]_{q}\,
z + q^{4} \frac{1}{[2]_{q}!}\Big[\begin{array}{c}3\\2\end{array}\Big]_{q}\, 
z^{2} - q^{9}\, \frac{1}{[3]_{q}!}\,z^{3},
\label{L_3}
\eeq
which, of course, agrees with the well-known \cite{M81, FV91} result. 
This 
validates our expansion scheme of the \qL\, polynomials in terms of the 
classical Laguerre polynomials. 
 
\section{$q$-Gegenbauer polynomials}
\label{q_Gegen}
\setcounter{equation}{0}
Classical Gegenbauer polynomials and their generalizations appear 
in many areas of theoretical physics, such as the correlation 
functions for the Kniznik-Zamolodchikov (KZ) \cite{KZ84} equation   
for the $\widehat{sl_{2}}$ algebra, and the wave functions of the  
integrable systems \cite{OP77} generalized from the Calogero-Sutherland
\cite{C71,S72} model. Deformed $q$-Gegenbauer polynomials 
have been introduced by Askey and Ismail \cite{AI83}. Naturally these 
deformed polynomials are of interest in studying the $q$-KZ equation 
for the $U_{q}(\widehat{sl_{2}})$ algebra \cite{FZ95}, and also the  
deformations of integrable models \cite{OS04}. We think our method of 
expressing 
$q$-Gegenbauer polynomials as nonlinear superpositions of classical 
Gegenbauer polynomials may be of use in these studies.    

\par

The generating function \cite{AI83} of the $q$-Gegenbauer polynomials 
read   
\be
\mathfrak {G}^{(\lambda)}_{q} (\cos \theta,\,t) & \equiv & 
\frac{(q^{\lambda}\,\exp (i\, \theta) \,t; q)_{\infty}}
{(\exp (i\, \theta) \,t; q)_{\infty}}\,\,
\frac{(q^{\lambda}\,\exp (- i\, \theta) \,t; q)_{\infty}}
{(\exp (- i\, \theta) \,t; q)_{\infty}}\nn\\
& = & \frac{E_{q}( - q^{\lambda}\,\exp (i\, \theta) \,t)}
{E_{q}(- \exp (i\, \theta) \,t)}\,\,
\frac{E_{q}(- q^{\lambda}\,\exp (- i\, \theta) \,t)}
{E_{q}(-\exp (- i\, \theta) \,t)}\nn\\
& = & \sum_{n = 0}^{\infty} C_{n}^{(\lambda)} (\cos \theta; q)\,t^{n}.
\label{q_G_gen}
\ee
The $q$-Gegenbauer polynomials are explicitly given by \cite{AI83}
\beq
C_{n}^{(\lambda)} (\cos \theta; q) =  \sum_{\ell = 0}^{n}\,
\frac{(q^{\lambda}; q)_{\ell}\,(q^{\lambda}; q)_{n - \ell}} 
{(q; q)_{\ell}\,(q; q)_{n - \ell}}\,\cos (n -2 \ell)\theta. 
\label{qG_lst}
\eeq  
In contrast to the case of Hermite and Laguerre polynomials discussed 
earlier the generating function for the classical Gegenbauer polynomials 
is not usually expressed in terms of ordinary exponentials:
\beq
\mathfrak {G}^{(\lambda)} (\cos \theta,\,t) \equiv 
(1 -2\,\cos \theta\,\,t + t^{2})^{- \lambda}    
= \sum_{n = 0}^{\infty} C_{n}^{(\lambda)} (\cos \theta)\,t^{n}.
\label{cl_G_gen}
\eeq
For subsequent applications we here list the classical Gegenbauer 
polynomials as follows:
\beq
C_{n}^{(\lambda)}(\cos \theta) = \sum_{\ell = 0}^{n}\,
\frac{(\lambda)_{\ell}\,(\lambda)_{n - \ell}}
{\ell!\,(n - \ell)!}\,\cos(n - 2\,\ell) \theta.
\label{cl_G_lst}
\eeq  
To establish a connection formula between the $q$-deformed Gegenbauer 
polynomials and their classical analogs we recast the classical 
generating function (\ref{cl_G_gen}) in an alternate form:
\beq
\mathfrak {G}^{(\lambda)} (\cos \theta,\,t) 
= \exp \Big(2\, \lambda \,\sum_{k \in \mathbb{N}} \cos\, (k \theta)\, 
\frac{{t^k}}{k}\Big).
\label{cl_G_exfrm}
\eeq
In a parallel construction we use the expansion scheme (\ref{Eqseries}) 
to express the deformed generating function 
$\mathfrak {G}^{(\lambda)}_{q} (\cos \theta,\,t)$ given in 
(\ref{q_G_gen}) as an infinite product series of the ordinary 
exponentials:  
\beq
\mathfrak {G}^{(\lambda)}_{q} (\cos \theta,\,t) 
= \exp \Big(2\, \sum_{k \in \mathbb{N}} 
\,[\lambda ]_{q^{k}}\,\,
\cos\, (k \theta)\, \frac{{t^k}}{k}\Big).
\label{q_G_exfrm}
\eeq
The close kinship between the product serieses (\ref{cl_G_exfrm})
and (\ref{q_G_exfrm}) now provides an interrelation between 
these two generating functions:  
\beq
\mathfrak {G}^{(\lambda)}_{q} (\cos \theta,\,t) 
= \exp \Big(\frac{1 - q^{\lambda {\cal D}}}{1 - q^{\cal D}}\,
\ln \mathfrak {G}^{(1)} (\cos \theta,\,t)\Big),
\label{gegen_G_rel}
\eeq
where ${\cal D} \equiv t\,\partial_{t}$. For the purpose of simplicity
we, in the above expression, have used the classical generating function 
for the $\lambda = 1$ case, namely 
$\mathfrak {G}^{(1)} (\cos \theta,\,t)$. In this case the 
Gegenbauer polynomials reduce to the Chebyshev polynomials of the 
second kind. 

\par

Starting from the mapping (\ref{gegen_G_rel}) of the deformed 
generating function on the classical generating function it is 
possible to develop via 
the route adopted in Secs. \ref{q_Laguerre} and \ref{q_Hermite} a general 
nonlinear connection formula between an arbitrary $q$-Gegenbauer 
polynomial and its classical partners. But since the general formula is 
notationally quite cumbersome, we subsequently express the first few 
$q$-Gegenbauer polynomials in terms of their classical analogs. But prior 
to that it is worthwhile to recast (\ref{gegen_G_rel}) in another form 
particularly suitable for deriving a set of {\it sum rules}:   
\beq
\ln \mathfrak {G}^{(\lambda)}_{q} (\cos \theta,\,t) 
= [\lambda]_{q^{\cal D}}\,\ln \mathfrak {G}^{(1)} (\cos \theta,\,t),
\label{G_sumrule}
\eeq
Expanding both sides of (\ref{G_sumrule}) in the variable $t$ and 
equating its identical powers, we obtain the general sum rule:
\be
&&\sum_{n \in \mathbb{N}} \frac{(- 1)^{n + 1}}{n}\,
\sum_{\ell_{1},\cdots,\ell_{n} \atop \in \mathbb{N}}\,
C_{\ell_{1}}^{(\lambda)}(z; q)\cdots
C_{\ell_{n}}^{(\lambda)}(z; q)\,\delta_{\ell_{1} + \cdots + \ell_{n},\, 
{\cal N}}\nn\\
&&= \,[\lambda]_{q^{\cal N}}\sum_{n \in \mathbb{N}} 
\frac{(- 1)^{n + 1}}{n}\,
\sum_{\ell_{1},\cdots,\ell_{n} \atop \in \mathbb{N}}\,
C_{\ell_{1}}(z)\cdots C_{\ell_{n}}(z)\,
\delta_{\ell_{1} + \cdots + \ell_{n},\, {\cal N}}.
\label{gegen_sum}
\ee
As mentioned in the context of (\ref{gegen_G_rel}), we here and 
henceforth consider the classical Gegenbauer polynomials 
$C_{n}^{(\lambda)}(z)$ for the $\lambda = 1$ case and suppress the 
superscript. For an explicit value of ${\cal N} \in \mathbb{N}$ the 
sum rule immediately follows from (\ref{gegen_sum}). 
To illustrate the first few 
cases we introduce, via (\ref{gegen_sum}), a set of variables:  
\be
{\cal I}_{1}^{(\lambda)}(z; q) &=& C_{1}^{(\lambda)}(z; q) \qquad 
{\cal I}_{2}^{(\lambda)}(z; q) \,=\, C_{2}^{(\lambda)}(z; q) - 
\frac{1}{2}\,(C_{1}^{(\lambda)}(z; q))^{2}\nn\\ 
{\cal I}_{3}^{(\lambda)}(z; q) &=&  C_{3}^{(\lambda)}(z; q) - 
C_{1}^{(\lambda)}(z; q)\,C_{2}^{(\lambda)}(z; q) 
- \frac{1}{3} (C_{1}^{(\lambda)}(z; q))^{3}\nn\\
{\cal I}_{4}^{\lambda)}(z; q) &=& C_{4}^{(\lambda)}(z; q) - 
C_{1}^{(\lambda)}(z; q)\,C_{3}^{(\lambda)}(z; q) 
- \frac{1}{2} (C_{2}^{(\lambda)}(z; q))^{2}\nn\\
& & + (C_{1}^{(\lambda)}(z; q))^{2}\,C_{2}^{(\lambda)}(z; q)
- \frac{1}{4} (C_{1}^{(\lambda)}(z; q))^{4}\nn\\
{\cal I}_{5}^{\lambda)}(z; q) &=& C_{5}^{(\lambda)}(z; q) - 
C_{1}^{(\lambda)}(z; q)\,C_{4}^{(\lambda)}(z; q) 
- C_{2}^{(\lambda)}(z; q)\,C_{3}^{(\lambda)}(z; q)\nn\\ 
& & + (C_{1}^{(\lambda)}(z; q))^{2}\,C_{3}^{(\lambda)}(z; q)
+ C_{1}^{(\lambda)}(z; q)\,(C_{2}^{(\lambda)}(z; q))^{2}\nn\\
& & - (C_{1}^{(\lambda)}(z; q))^{3}\,C_{2}^{(\lambda)}(z; q)
+ \frac{1}{5} (C_{1}^{(\lambda)}(z; q))^{5}.
\label{nonli_comb}
\ee
The sum-rules given in (\ref{gegen_sum}) may then be succinctly 
stated as
\beq
{\cal I}_{\ell}^{(\lambda)}(z; q) = [\lambda]_{q^{\ell}}
\big({\cal I}_{\ell}^{(\lambda = 1)}(z; q)\big)_{q \rightarrow 1} \quad 
\forall 
\ell \in {\mathbb N}
\label{suc_sum}
\eeq
We also enumerate a few explicit examples of our nonlinear connection 
formula relating $q$-Gegenbauer polynomials and their classical 
counterparts:  
\be
C_{0}^{(\lambda)}(z; q) &=& C_{0}(z)\; = \;1,\qquad
C_{1}^{(\lambda)}(z; q)\; = \;[\lambda]_{q} C_{1}(z),\nn\\
C_{2}^{(\lambda)}(z; q) &=& [\lambda]_{q^{2}} C_{2}(z) - 
\frac{1}{2} ([\lambda]_{q^{2}} - [\lambda]_{q}^{2})\, C_{1}^{2}(z),\nn\\
C_{3}^{(\lambda)}(z; q) &=& [\lambda]_{q^{3}} C_{3}(z) - 
([\lambda]_{q^{3}} - [\lambda]_{q}\,[\lambda]_{q^{2}})
\,C_{1}(z)\, C_{2}(z)\nn\\
& & + \frac{1}{6}\,(2\,[\lambda]_{q^{3}}
- 3\,[\lambda]_{q}\,[\lambda]_{q^{2}} + [\lambda]_{q}^{3})\,
\,C_{1}^{3}(z),\nn\\
C_{4}^{(\lambda)}(z; q) &=& [\lambda]_{q^{4}} C_{4}(z)  
- \frac{1}{2} ([\lambda]_{q^{4}} - [\lambda]_{q^{2}}^{2})\, 
C_{2}^{2}(z)\nn\\
& & - ([\lambda]_{q^{4}} - [\lambda]_{q}\,[\lambda]_{q^{3}})
\,C_{1}(z)\, C_{3}(z)\nn\\
& & + \frac{1}{2}\,(2\,[\lambda]_{q^{4}}
- 2\,[\lambda]_{q}\,[\lambda]_{q^{3}} - 
[\lambda]_{q^{2}}^{2} + [\lambda]_{q}^{2}\,[\lambda]_{q^{2}}) 
\,C_{1}^{2}(z)\, C_{2}(z)\nn\\
& & - \frac{1}{24}\,(6\,[\lambda]_{q^{4}} - 
8\,[\lambda]_{q}\,[\lambda]_{q^{3}} - 3\,[\lambda]_{q^{2}}^{2} + 6\, 
[\lambda]_{q}^{2}\,[\lambda]_{q^{2}} - [\lambda]_{q}^{4})\, 
C_{1}^{4}(z),\nn\\
C_{5}^{(\lambda)}(z; q) &=& [\lambda]_{q^{5}} C_{5}(z)  
- ([\lambda]_{q^{5}} - [\lambda]_{q}\,[\lambda]_{q^{4}})\, 
C_{1}(z)\,C_{4}(z)\nn\\
& & - ([\lambda]_{q^{5}} - [\lambda]_{q^{2}}\,[\lambda]_{q^{3}})
\,C_{2}(z)\, C_{3}(z)\nn\\
& & + \frac{1}{2}\,(2\,[\lambda]_{q^{5}}- 2\,[\lambda]_{q}\,
[\lambda]_{q^{4}} - [\lambda]_{q^{2}}\, [\lambda]_{q^{3}} 
+ [\lambda]_{q}^{2}\, [\lambda]_{q^{3}}) 
\,C_{1}^{2}(z)\, C_{3}(z)\nn\\
& & + \frac{1}{2}\,(2\,[\lambda]_{q^{5}}- [\lambda]_{q}\,
[\lambda]_{q^{4}} - 2 [\lambda]_{q^{2}}\, [\lambda]_{q^{3}} 
+ [\lambda]_{q}\, [\lambda]_{q^{2}}^{2}) 
\,C_{1}(z)\, C_{2}^{2}(z)\nn\\
& & - \frac{1}{6}\,(6 \,[\lambda]_{q^{5}}- 6 \,[\lambda]_{q}\,
[\lambda]_{q^{4}} - 5 [\lambda]_{q^{2}}\, [\lambda]_{q^{3}} 
+ 3 \,[\lambda]_{q}^{2}\, [\lambda]_{q^{3}} 
+ 3 \,[\lambda]_{q} \,[\lambda]_{q^{2}}^{2}\nn\\ 
& & - [\lambda]_{q}^{3} \,[\lambda]_{q^{2}}) 
\,C_{1}^{3}(z)\, C_{2}(z)\nn\\
& & + \frac{1}{120}\,(24\,[\lambda]_{q^{5}} 
- 30\,[\lambda]_{q}\,[\lambda]_{q^{4}}  
- 20\,[\lambda]_{q^{2}}\,[\lambda]_{q^{3}} 
+ 20 \,[\lambda]_{q}^{2} \,[\lambda]_{q^{3}}\nn\\  
& & + 15\,[\lambda]_{q}\,[\lambda]_{q^{2}}^{2} 
- 10\,[\lambda]_{q}^{3}\,[\lambda]_{q^{2}}
+ [\lambda]_{q}^{5})\, 
C_{1}^{5}(z).
\label{expl_qG}
\ee
Explicit use of the above connection formulae immediately proves the 
validity of the sum rules presented in (\ref{suc_sum}). It is 
interesting to note that modulo a multiplicative factor, the 
combinations ${\cal I}_{\ell}^{(\lambda)}(z; q)$ preserve their `form'    
when the deformed polynomials are recast in terms of their 
classical partners via our connection formulae. 
 
\section{Discussions}
Using a realization of the $q$-exponential function as an infinite 
multiplicative series of the ordinary exponentials we have obtained a 
new set of nonlinear connection formulae for the $q$-orthogonal 
polynomials in terms of their classical partners. The scheme has been 
illustrated for $q$-Hermite, $q$-Laguerre and $q$-Gegenbauer 
polynomials. Our procedure may be applied for other instances such as 
little $q$-Jacobi polynomials. The matrix elements of the finite 
dimensional unitary co-representations of the quantum group 
$SU_{q}(2)$ and the supergroup $OSp_{q}(1/2)$ may be obtained 
\cite{MM90, ACNS06} using the little $q$-Jacobi polynomials. Our 
formulation allows expressing the said matrix elements via 
their classical analogs. It may, for instance, be useful in constructing 
so far unknown generating function \cite{N91} of the 
co-representation matrices of the quantum supergroup $OSp_{q}(1/2)$.
We will address this issue elsewhere.    

\section*{Acknowledgement}
Two of us (R.C. and S.S.N.M.) are partially supported by the grant
DAE/2001/37/12/BRNS, Government of India. 

\newpage
%%%
\clearpage
\begin{table}[!ht]
{\small
\begin{center}
\begin{tabular}{c||c} 
\hline Solutions of (\ref{Her_part}) for $n = 5$ & 
Contributions to $H_{5}(z;q)$ \\ 
\hline
$n_{1} = 5$ & $ -\frac{\5}{5!}\,
\Big(\frac{2}{q^{2}\, [2]_{q^{-2}}}\Big)^{5/2}\,H_{5}(\zeta_{1})$\\ 
\hline
$n_{1} = 3, n_{2} = 1$ & $\sqrt{- 1}\,\,\frac{\5}{3!}\, 
\Big(\frac{2}{q^{2}\, [2]_{q^{-2}}}\Big)^{5/2}\,\sqrt{c_{2}(q^{-4})}\,
H_{3}(\zeta_{1}) \, H_{1}(\zeta_{2})$\\ \hline
$n_{1} = 2, n_{3} = 1$ & $ \frac{\5}{2!}
\Big(\frac{2}{q^{2}\, [2]_{q^{-2}}}\Big)^{5/2}\,\sqrt{c_{3}(q^{-4})}\,
H_{2}(\zeta_{1}) \, H_{1}(\zeta_{3})$\\ \hline
$n_{1} = 1, n_{4} = 1$ & $ - \sqrt{-1}\,\5\,
\Big(\frac{2}{q^{2}\, [2]_{q^{-2}}}\Big)^{5/2}\,\sqrt{c_{4}(q^{-4})}\,
H_{1}(\zeta_{1}) \, H_{1}(\zeta_{4})$\\ \hline
$n_{2} = 1, n_{3} = 1$ & $ - \sqrt{-1}\,\5\,
\Big(\frac{2}{q^{2}\, [2]_{q^{-2}}}\Big)^{5/2}\,
\sqrt{c_{2}(q^{-4})\, c_{3}(q^{-4}))}\,
H_{1}(\zeta_{2}) \, H_{1}(\zeta_{3})$\\ \hline
$n_{1} = 1, n_{2} = 2$ & $\frac{\5}{2!}\,
\Big(\frac{2}{q^{2}\, [2]_{q^{-2}}}\Big)^{5/2}\,c_{2}(q^{-4})\, 
H_{1}(\zeta_{1}) \, H_{2}(\zeta_{2})$\\ \hline
$n_{5} = 1$ & $- \5 \,\,\Big(\frac{2}{q^{2}\, [2]_{q^{-2}}}\Big)^{5/2}\,
\sqrt{c_{5}(q^{-4})}\, H_{1}(\zeta_{5})$\\ \hline
\end{tabular}
\caption{Contributions to $H_{5}(z;q)$}
\label{H_T}
\end{center}
}
\end{table}

\clearpage

\begin{table}[!ht]
{\small
\begin{center}
\begin{tabular}{c||c} 
\hline Solutions of (\ref{part_eq}) & 
Contributions to $L_{3}^{(0)}(z;q)$ \\ \hline
$k_{1} = 3$ & $q^{6} L^{(n_{1} - 3)}_{3}(z)$\\ \hline   
$\ell_{1} = 3$ & $- q^{6}\,\frac{(n_{1})_{3}}{3 !}$\\ \hline
$\ell = 3$ & $1$\\ \hline 
$k_{3} = 1$ &
$q^{6}\, L_{1}^{(n_{3} - 1)}(c_{3}(q)\,z^{3})$\\ \hline
$\ell_{3} = 1$ & $- q^{6} n_{3}$\\ \hline 
$k_{1} = 1, k_{2} = 1$ &
$q^{6} L_{1}^{(n_{1} - 1)}(z)\, 
L_{1}^{(n_{2} - 1)}(c_{2}(q)\,z^{2})$\\ \hline 
$\ell_{1} = 1, \ell_{2} = 1$ & $q^{6} n_{1} n_{2}$\\ \hline
$k_{1} = 2, \ell_{1} = 1$ & 
$- q^{6} n_{1} L_{2}^{(n_{1} - 2)}(z)$\\ \hline 
$k_{1} = 1, \ell_{1} = 2$ &
$q^{6} \frac{(n_{1})_{2}}{2} \, L_{1}^{(n_{1} - 1)}(z)$\\ \hline 
$k_{1} = 2, \ell = 1$ & 
$q^{3}\,[3]_{q}\,L_{2}^{(n_{1} - 2)}(z)$\\ \hline 
$k_{1} = 1, \ell = 2$ &
$q \, [3]_{q}\,L_{1}^{(n_{1} - 1)}(z)$\\ \hline 
$\ell = 1, \ell_{1} = 2$ & 
$q^{3} \, [3]_{q}\, \frac{(n_{1})_{2}}{2}$\\ \hline  
$\ell = 2, \ell_{1} = 1$ &
$- q \, [3]_{q}\, n_{1}$\\ \hline 
$k_{2} = 1, \ell_{1} = 1$ &
$- q^{6} n_{1} \,L_{1}^{(n_{2} - 1)}(c_{2}(q)\,z^{2})$\\ \hline  
$k_{2} = 1, \ell = 1$ &
$q^{3}\, [3]_{q} \,L_{1}^{(n_{2} - 1)}(c_{2}(q)\,z^{2})$\\ \hline
$\ell = 1, \ell_{2} = 1$ & $- q^{3} \, [3]_{q}\, n_{2}$\\ \hline
$k_{1} = 1, \ell_{2} = 1$ & 
$- q^{6} n_{2} \,L_{1}^{(n_{1} - 1)}(z)$\\ \hline 
$k_{1} = 1, \ell = 1, \ell_{1} = 1$ &
$-q^{3} \, [3]_{q}\, n_{1} \,L_{1}^{(n_{1} - 1)}(z)$\\ \hline
\end{tabular}
\caption{Contributions to $L_{3}^{(0)}(z;q)$}
\label{L_T}
\end{center}
}   
\end{table}

\newpage

\suppressfloats

\suppressfloats

\end{document}